\documentclass[aps,prl,twocolumn,amsmath,amssymb,superscriptaddress,floatfix]{revtex4-1}

\usepackage{amsmath}
\usepackage{amssymb}
\usepackage{amsfonts}
\usepackage{graphicx}
\usepackage{bm}
\usepackage[colorlinks=true,linkcolor=blue,citecolor=blue,urlcolor=blue]{hyperref}

\newcommand{\SSD}{\mathrm{SSD}}
\newcommand{\HH}{\mathcal{H}}
\newcommand{\dd}{\mathrm{d}}
\newcommand{\ee}{\mathrm{e}}
\newcommand{\ii}{\mathrm{i}}

\begin{document}

\title{Emergent Black Hole Dynamics in Critical Floquet Systems}

\author{Bastien Lapierre}
\affiliation{Institute for Theoretical Physics,	ETH Z\"urich,	Wolfgang-Pauli-Str. 27, 8093 Z\"urich, Switzerland}
\affiliation{Department of Physics, University of Z\"urich, Winterthurerstrasse 190, 8057 Z\"urich, Switzerland}

\author{Kenny~Choo} 
\affiliation{Department of Physics, University of Z\"urich, Winterthurerstrasse 190, 8057 Z\"urich, Switzerland}

\author{Cl\'ement Tauber}
\affiliation{Institute for Theoretical Physics,	ETH Z\"urich,	Wolfgang-Pauli-Str. 27, 8093 Z\"urich, Switzerland}

\author{Apoorv Tiwari}
\affiliation{Department of Physics, University of Z\"urich, Winterthurerstrasse 190, 8057 Z\"urich, Switzerland}
\affiliation{Condensed Matter Theory Group, Paul Scherrer Institute, CH-5232 Villigen PSI, Switzerland}

\author{Titus Neupert}
\affiliation{Department of Physics, University of Z\"urich, Winterthurerstrasse 190, 8057 Z\"urich, Switzerland}

\author{Ramasubramanian Chitra}
\affiliation{Institute for Theoretical Physics,	ETH Z\"urich,	Wolfgang-Pauli-Str. 27, 8093 Z\"urich, Switzerland}

\date{\today}

\begin{abstract}
While driven interacting quantum matter is generically subject to heating and scrambling, certain classes of systems evade this paradigm.  
We study such an exceptional class in periodically driven critical $(1+1)$-dimensional systems with a spatially modulated, but disorder-free time evolution operator.
Instead of complete scrambling, the excitations of the system remain well-defined. Their propagation is analogous to the evolution along light cones in a curved space-time obtained  by two Schwarzschild black holes. The Hawking temperature 
serves as an order parameter which distinguishes between heating and non-heating phases. Beyond a time scale
determined by the inverse Hawking temperature, excitations are
absorbed by the black holes resulting in a singular concentration of energy at their center.
We obtain these results analytically within conformal field theory, capitalizing on a mapping to sine-square deformed field theories. Furthermore, by means of numerical calculations for an interacting XXZ spin-$\frac 12$ chain, we demonstrate that our findings survive lattice regularization.

\end{abstract}

\maketitle

\textit{Introduction} ---
Floquet quantum many-body systems  provide a rich arena to
explore new foundational principles of statistical physics beyond equilibrium.  Interest in the field is further spurred on by experimental advances  in quantum engineered systems which permit the exploration of Floquet physics in a controlled manner \cite{PhysRevX.8.021030,PhysRevLett.119.200402}.
The prevailing paradigm  in driven systems is   that  closed integrable systems converge to steady states described by generalized Gibbs ensembles while  interacting Floquet systems heat up to trivial infinite temperature states where
all notion of coherence is lost \cite{PhysRevLett.112.150401}.    
Generically,  whether a system heats up or not is intimately linked  to   notions of integrability, ergodicity, and quantum chaos.  
 A deeper understanding  of how a system heats  up due  the interplay between drive and interactions  is still lacking.
This is attributable to the limited scope of  analytical  and numerical  methods available to study such complex many body systems. 
 
A pioneering effort in this direction was the recently  proposed  Floquet  conformal field theory (CFT)~\cite{Wen:2018vux,Wen:2018agb},   based on  a sine square deformation (SSD) of a CFT.    SSDs  impose a  specifc inhomogeneous energy density profile, and were originally proposed  as a numerical  trick for quantum simulations~\cite{Gendiar_2009,Toshiya_2011, Maruyama_2011}.  The richness of SSDs was only recently unveiled via studies of   both integrable lattice models like quantum spin chains~\cite{Katsura_2012, Allegra_2016, Dubail_2017} and CFTs in the continuum~\cite{ Ishibashi_2015, Okunishi_2016, Wen_2016, Tamura_2017, Wen:2018vux,Wen:2018agb}. 
It was exploited in Refs.~\onlinecite{Wen:2018vux,Wen:2018agb}, in which a  Floquet drive alternating between a  generic CFT and its SSD analogue was studied. Analyzing the entanglement entropy,  an intriguing phase diagram with transitions between heating and non-heating phases was obtained.

In this letter,  we  address the question, {\it how does a system heat up ?}  in this class of systems via an analytical  study of  dynamical two-point correlation functions  in the CFT formalism  and a parallel  numerical study verifying the  surprising robustness of the CFT predictions  in a genuine one-dimensional  XXZ quantum spin chain at criticality subject to the Floquet-SSD protocol. 
By analyzing the time-evolution of two-point correlation functions, we show that the heating phase is characterized by the emergence of  stroboscopic black hole singularities, which manifest  as attractors at two spatial locations towards which all  excitations  evolve  and where energy  accumulates indefinitely,  irrespective of initial conditions.  
\begin{figure}[htb]
	\includegraphics[width=8.6cm]{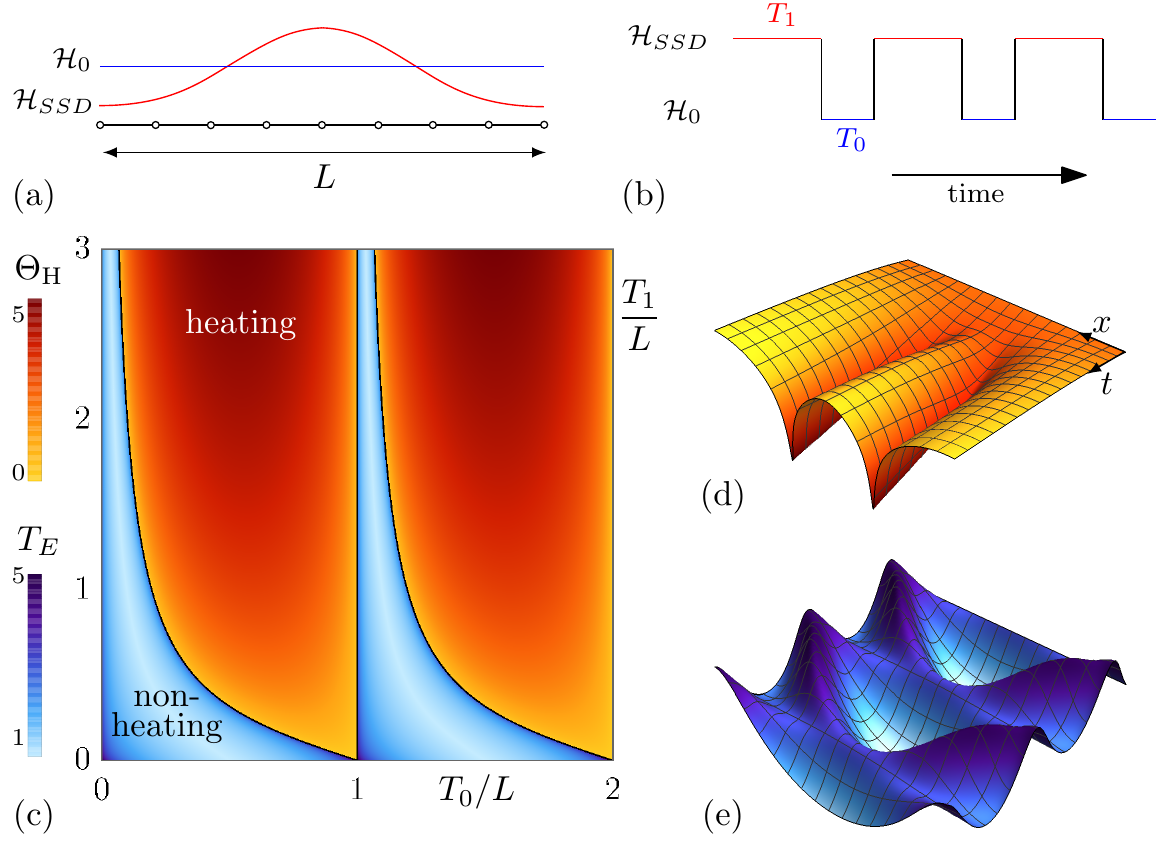}
	\caption{\label{fig:intro} (a) Uniform and SSD Hamiltonian. (b) Floquet drive alternating between the two. (c) Phase diagram (colorbars in log scale). The heating phase is characterized by a Hawking temperature $\Theta_\mathrm{H}$, a signature of emergent black holes in the effective dynamics, whose space-time is illustrated in (d). The non-heating phase is characterized by a pseudo-periodicity $T_E$, with  effective space-time  illustrated in (e). The phase diagram is $\tfrac{T_0}{L}$ periodic.}
\end{figure}
We show that the   associated Hawking temperature of the black holes  serves as  {\it de facto} order parameter which
delineates heating and non-heating phases.  The non-heating phase manifests a pseudo-periodicity both in the propagation of excitations as well as energy density.

\textit{Floquet-SSD Dynamics} --- Consider a general inhomogeneous Hamiltonian on a chain of size $L$ obtained by deforming a uniform $(1+1)$-dimensional CFT:
 \begin{equation}
	\HH = \int_0^L \dd x f(x) T_{00}(x).
\end{equation}
We denote by $\HH_0$ the homogeneous CFT where $f\equiv 1$, with energy density $T_{00}$, and by $\HH_\SSD$ the SSD theory where $f(x) = 2\sin^2(\tfrac{\pi x}{L})$. We consider  a two-step   drive  protocol, where $\HH_\mathrm{F}(t)$  alternates between   $\HH_\SSD$ (duration $T_1$)   and  $\HH_0$ (duration $T_0$)  as depicted in
  Fig.~\ref{fig:intro}(a,b).
  The uniform theory $\HH_0$  typically  describes the low-energy behavior of a quantum chain at criticality and is characterized by a central charge $c$.

 The  lattice  counterpart we  explicitly consider  is the  XXZ spin-$\frac 12$  chain,
\begin{equation}\label{spinchain}
	H = J \sum_{j=1}^{L-1} f_j \Big( S_j^x S_{j+1}^x + S_j^y S_{j+1}^y + \Delta S_j^z S_{j+1}^z \Big),
\end{equation}
The Floquet drive $H_\mathrm{F}(t)$ alternates  between the
the uniform case, $H_0$ with   $f_j \equiv 1$ and the SSD $H_\SSD$ where $f_j = 2\sin^2(\tfrac{\pi j}{L})$.
  For   $f_j\equiv 1$ and  $|\Delta|\le1$, the spin chain is critical  and the low energy theory is a  Luttinger liquid described by a compactified free boson with $c=1$.
  In what follows, we will demonstrate that
 the general non-equilibrium exactly solvable CFT-dynamics of $\HH_\mathrm{F}(t)$  precisely captures the main features of the driven XXZ model $H_\mathrm{F}(t)$ that we study numerically.

To probe the dynamics we focus on the unequal time  two-point function of the driven CFT $\HH_\mathrm{F}(t)$, $\langle \phi(x,t) \phi(x_0,0)\rangle$, where $\phi$ is any primary field (with conformal weight $h$) of the uniform theory $\HH_0$ \cite{Dubail_2017}.  Though the full time evolution including micromotion can be evaluated, we focus on the stroboscopic evolution, where $t=n(T_0+T_1)$, $n\in \mathbb{N}$.  As boundary conditions  do not qualitatively affect the ensuing results,   we use periodic boundary conditions for computational simplicity. Expectation values are computed  in the ground state $|0\rangle$ of the uniform theory.  In  terms of the Virasoro generators $L_n$ and $\overline L_n$,  in the Euclidean framework with imaginary time $\tau$,  
 $\HH_0 = L_0 + \overline L_0$,   and crucially, $\HH_\SSD = L_0 - \tfrac{1}{2}(L_1 + L_{-1}) + \overline L_0 - \tfrac{1}{2}(\overline L_1 + \overline L_{-1})$.  Such a Hamiltonian is equivalent to a uniform $\mathcal H_{0}$ up to an asymptotic $SL(2,\mathbb R)$ transformation \cite{Ishibashi_2015, Okunishi_2016, Wen:2018vux}. Consequently, time evolution $\ee^{-\tau {\HH_\SSD}}$ is a simple dilation up to a coordinate change. Mapping  the  coordinates $w = \tau + \ii x$  on the cylinder to the complex plane spanned by $z=\ee^{2\pi w/L}$,  the CFT  calculation yields, after analytic continuation to real time~(see Supplemental Material for details~\cite{supmat})
 \begin{equation}\label{eq:two_point_CFT}
 	\langle \phi(x,t) \phi(x_0,0) \rangle = \left[ \Big(\frac{2\pi}{L}\Big)^{\hspace{-0.1cm} 4} \dfrac{\partial \tilde{z}_n}{\partial z} \dfrac{\partial \bar{\tilde z}_n}{\partial \bar z}  \right]^{\hspace{-0.cm}h}\hspace{-0.17cm}\langle \phi(\tilde{z}_n, \bar{\tilde{z}}_n) \phi(\tilde z_0, \bar{\tilde z}_0) \rangle,
 \end{equation}
 where the two-point function on the right hand side corresponds to  the one evaluated in the uniform CFT, namely $\langle \phi(\tilde{z}_n, \bar{\tilde z}_n) \phi(\tilde z_0, \bar{\tilde z}_0) \rangle = (\tilde{z}_n-\tilde z_0)^{-2h}(\bar{\tilde z}_n-\bar{\tilde z}_0)^{-2h}$. Remarkably,  the  nontrivial   Floquet dynamics  is fully encoded in the change of variables, which is essentially a  M\"obius transformation \cite{Wen:2018agb}
\begin{equation}\label{eq:tildez}
	\tilde{z}_n = \dfrac{(\gamma_1 -\eta^n \gamma_2) z + (\eta^n-1)\gamma_1 \gamma_2}{(1-\eta^n)z + \gamma_1 \eta^n -\gamma_2}
\end{equation}
where  $n$ is  the number of drive cycles, $\eta, \gamma_1, \gamma_2$ are complex parameters that depend on $\tfrac{T_0}{L}$ and $\tfrac{T_1}{L}$~\cite{supmat}.   This result is valid for a generic CFT. The central charge enters only via the conformal dimensions of the operators in the correlation function.  { Moreover, although Eq.~\eqref{eq:two_point_CFT} captures the stroboscopic dynamics of $\HH_\mathrm{F}(t)$, it is well defined not only at discrete, but all continuous times -- a fact that we will exploit below \cite{supmat}.}  We now discuss the two distinct  regimes of behavior  classified by the  parameter $\eta$: 
(i) heating phase for  $\eta \in \mathbb R^+$
(ii) a non-heating phase with $\eta \in \mathbb C, |\eta |=1 $, with $\eta =1$ signalling the transition between the two. The corresponding phase diagram is given in Fig.~\ref{fig:intro}~(c).

 \begin{figure}[t]
	\includegraphics[width=8.6cm]{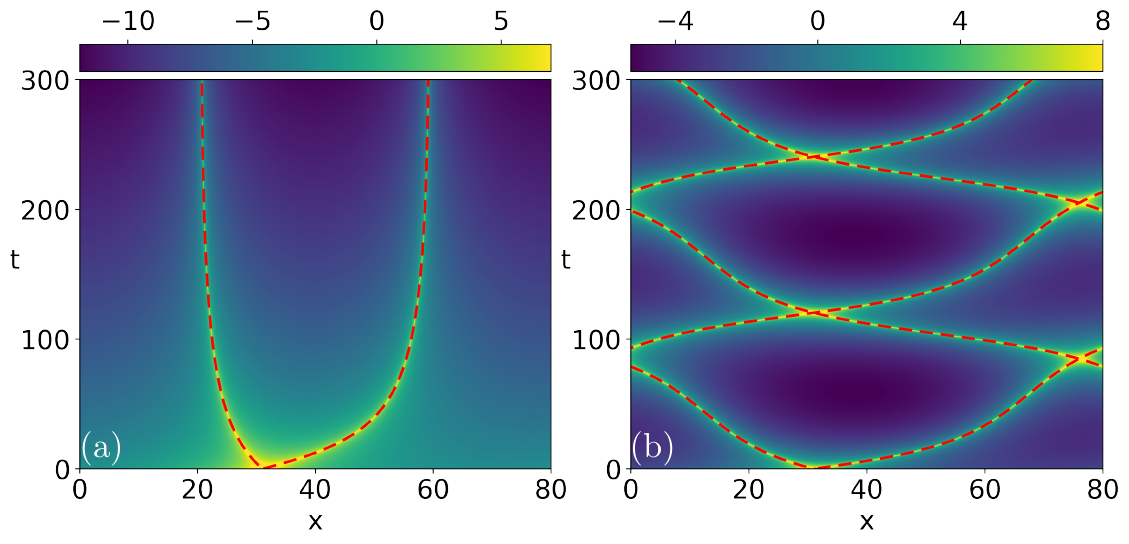}
	\caption{\label{fig:2pointCFT} CFT two-point function $|\langle \phi(x,t) \phi(x_0,0)\rangle|$ for the Floquet drive  ($L=80$, $x_0=31$, colorbars in log scale). (a) Heating phase ($T_0=T_1=34$). The excitations are attracted by two black hole singularities at $x_c$ and $L-x_c$. (b) Non-heating phase ($T_0=T_1=25$). The dynamics is pseudo-periodic. In both cases, the dashed curves are the null-geodesics of the curved stationary metric.}
\end{figure}

\textit{Heating phase.} --- In this regime, $\eta \in \mathbb R^+$, in which case $\tilde{z}_n\to \gamma_1$ or $\gamma_2$ as $n \to \infty$ (depending on the sign of $\eta -1$). A typical two-point function is plotted in Fig.~\ref{fig:2pointCFT}~(a). After an initial transient regime during which excitations move quasi-ballistically but start to lose their coherence,  the  correlation function  tends to  aggregate at two spatial locations: $x_c$ and $L-x_c$, independent of the initial condition $x_0$.  Furthermore, the magnitude of the two point function grows with time indicating that the excitations accumulate indefinitely at the two `horizons'. This phenomenon can be interpreted in analogy to black holes emerging from a curved space-time: The two-point function~\eqref{eq:two_point_CFT} is reduced to a uniform CFT computation in the variable $\tilde{z}_n$, for which the metric is flat, namely,  $\dd s^2 = \dd \tilde{u}_n^2 + \dd \tilde{v}_n^2$  where $\tilde{z}_n = \tilde{u}_n + \ii \tilde{v}_n$. The change of coordinates $\tilde{u}_n(x,\tau)$, $\tilde{v}_n(x,\tau)$ can be computed explicitly from Eq.~\eqref{eq:tildez}, so we may read-off the corresponding curved metric in the $(x,\tau)$-coordinates. Up to a Weyl transformation and after analytic continuation to real time, one obtains 
\begin{equation}\label{eq:metric}
	\dd s^2 = \dd x^2 - g(x)\dd t^2 + 2 h(x) \dd x \dd t,
\end{equation}
in the original coordinates~\cite{supmat}. Here, $g$ and $h$ are time-independent real functions. Therefore, the Floquet drive is equivalent to a free propagation in a stationary curved space-time. The null geodesics $\dd s^2=0$ for such a metric lead to a non-uniform velocity $v(x) = h(x) \pm \sqrt{h(x)^2+g(x)}$ and the corresponding trajectories fit perfectly with the analytic two-point function of Fig.~\ref{fig:2pointCFT}~(a). In  the  heating phase,  $v(x_c) = v(L-x_c) =0$ such that once excitations arrive at these locations, they are permanently trapped.

To explore further the analogy with black holes in this curved space time, we assume  that $h(x)=0$, valid for a time-reversal symmetric  driving protocol such
the  $\dd x \dd t$ term in Eq.~\eqref{eq:metric}  vanishes.  Consequently, $v(x)=\pm \sqrt{g(x)}$ with $g(x)>0$ and $g(x_c) = g(L-x_c) =0$. One infers $x_c = \tfrac{L}{2\pi} \arccos(\cos \tfrac{\pi T_0}{L} + \tfrac{L}{\pi T_1}  \sin \tfrac{\pi T_0}{L})$ and 
\begin{equation}\label{eq:g}
	v(x) = 2A \,\dfrac{\sin \big[\frac{\pi}{L} (x-x_c)\big]\sin\big[ \frac{\pi}{L} (x+x_c) \big] }{\cos \big(\frac{2 \pi x_c}{L}\big)  },
\end{equation}
with $A= \tfrac{1+ \gamma_1^2}{\gamma_1^2-1}\tfrac{L \log \eta}{2\pi \ii (T_0+T_1)} \in \mathbb{R}$ \cite{supmat}. 
Near the horizon $x_c$, one finds that at leading order in $x-x_c$,  $\dd s^2 = - \Theta_\mathrm{H}^2 (x-x_c)^2 \dd t^2 + \dd x^2$.  This is a Rindler metric which is equivalent to a  Schwarzschild  metric in $(1+1)$ dimensions, up to a coordinate change \cite{supmat}. The corresponding Hawking temperature 
\begin{equation}\label{eq:hawking}
\Theta_\mathrm{H}=\frac{|\log{(\eta)}|}{2\pi(T_0+T_1)}
\end{equation}
is  plotted in Fig.~\ref{fig:intro}~(c). The inverse Hawking temperature  provides a timescale after which the  excitations  are fully trapped at $x=x_c$.  This timescale diverges at the transition to the non-heating phase, where $\Theta_\mathrm{H}\to 0$.  A similar expansion of $f$ near $x=L-x_c$ leads to a similar metric with the same $\Theta_\mathrm{H}$.

\textit{Non-heating phase} --- A  typical two-point function in the  non-heating phase  is plotted in Fig.~\ref{fig:2pointCFT}~(b).  As expected, the excitations 
are coherent,  
resulting in an oscillatory behavior  characterized by a new periodicity $T_E = 2\pi \tfrac{T_0+T_1}{|\log(\eta)|}$,  reminiscent of discrete time crystals \cite{2018PhT....71i..32G,PhysRevA.92.023815}, see Fig.~\ref{fig:intro}~(c). 
Notice that $T_E$ is in general  not an integer  multiple of $T_0+T_1$ so it is only a pseudo-periodicity for the original dynamics.  Nevertheless, for some specific values of $T_0$ and $T_1$ such that $|\log(\eta)| = 2\pi p$ with $p \in \mathbb N$, $T_E$ is indeed an integer multiple of the underlying periodicity of the drive.  

The curved space-time interpretation presented earlier  is also valid in the non-heating phase i.e., the excitations move ballistically in curved space-time
 with the stationary metric \eqref{eq:metric}. The main difference is that  no black hole singularities exist and   the corresponding velocity $v$ for the null geodesics is nonzero everywhere, so the excitation can  traverse  the  entire physical extent of the system.   Our analytic results are well described by these geodesics. 

 \begin{figure}[t]
	\includegraphics[width=8.6cm]{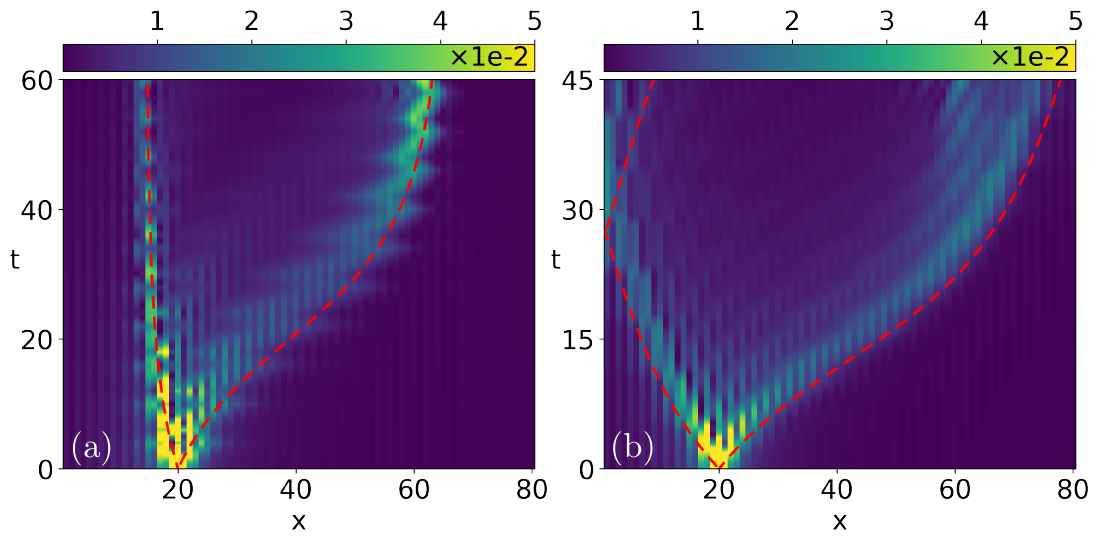}
	\caption{\label{fig:2pointXXZ} XXZ two-point function $|\langle S_z(x,t) S_z(x_0,0)\rangle |$ for the Floquet drive ($\Delta = 0.5$, $L=80$). (a) Heating phase ($T_0=2$, $T_1=4$) (b) Non-heating phase ($T_0=-2$, $T_1=4$). The dotted curves are the null-geodesics of the curved stationary metric from the CFT approach. }
\end{figure}

\textit{Driven XXZ model} --- To  see if  the emergent black holes in the  CFT analysis  survive in a realistic condensed matter setting, we  numerically simulate the  XXZ  spin  chain described in Eq.~\eqref{spinchain} subject to the Floquet-SSD driving protocol.   We use Matrix Product States (MPS) techniques to compute the two-point spin correlation function $\langle G| S^z(x,t) S^z(x_0,0)|G \rangle = \langle G| U^{\dagger}(t) S^z(x,0)U(t)S^z(x_0,0)|G \rangle$   both in the heating and non-heating phase. This is done using the ITensor library~\cite{ITensor} by simply taking the overlap between $U(t)S^z(x_0,0)|G \rangle$ and  $S^z(x,0)U(t)|G \rangle$, where the unitary evolution is implemented  by the sequential application of trotter gates. The numerically computed spin correlation functions are plotted in Fig.~\ref{fig:2pointXXZ} and manifest a remarkably good agreement with the  stroboscopic CFT predictions for  $c=1$   and a compactification radius {$R^2=\tfrac{2(\pi-\arccos\Delta)}{\pi}$.  $S^z$ corresponds to a combination of primary fields in the CFT sense}. In the heating phase,  we clearly see the emergence of the two predicted  black-holes singularities and the excitations follow the null geodesics of the curved space time \eqref{eq:metric}. Notice that, for numerical purposes we evolve with $-H_{0}$ instead of $H_0$, or equivalently we use a negative time $T_0$, which is fine because the phase diagram of Fig.~\ref{fig:intro} is $\tfrac{T_0}{L}$ periodic. In the non-heating phase, the black hole singularities are absent and the geodesics from CFT fit the numerical data well.   The  effective periodicity  $T_E$ is not seen in the simulations  due to computational limitations  on   long time  physics.

We briefly discuss the role of micromotion within a period.  We find that  micromotion  can indeed  lead to  additional interesting  features  both in the CFT as well as the  physical system on a lattice. Here, we focused on  regimes where  micromotion is reduced to small fluctuations around the stroboscopic dynamics, and can be neglected to first order.  A thorough  study of  the role of micromotion will be addressed in  future work \cite{CLTTNR-unpub}.  

\begin{figure}[t]
	\includegraphics[width=8.6cm]{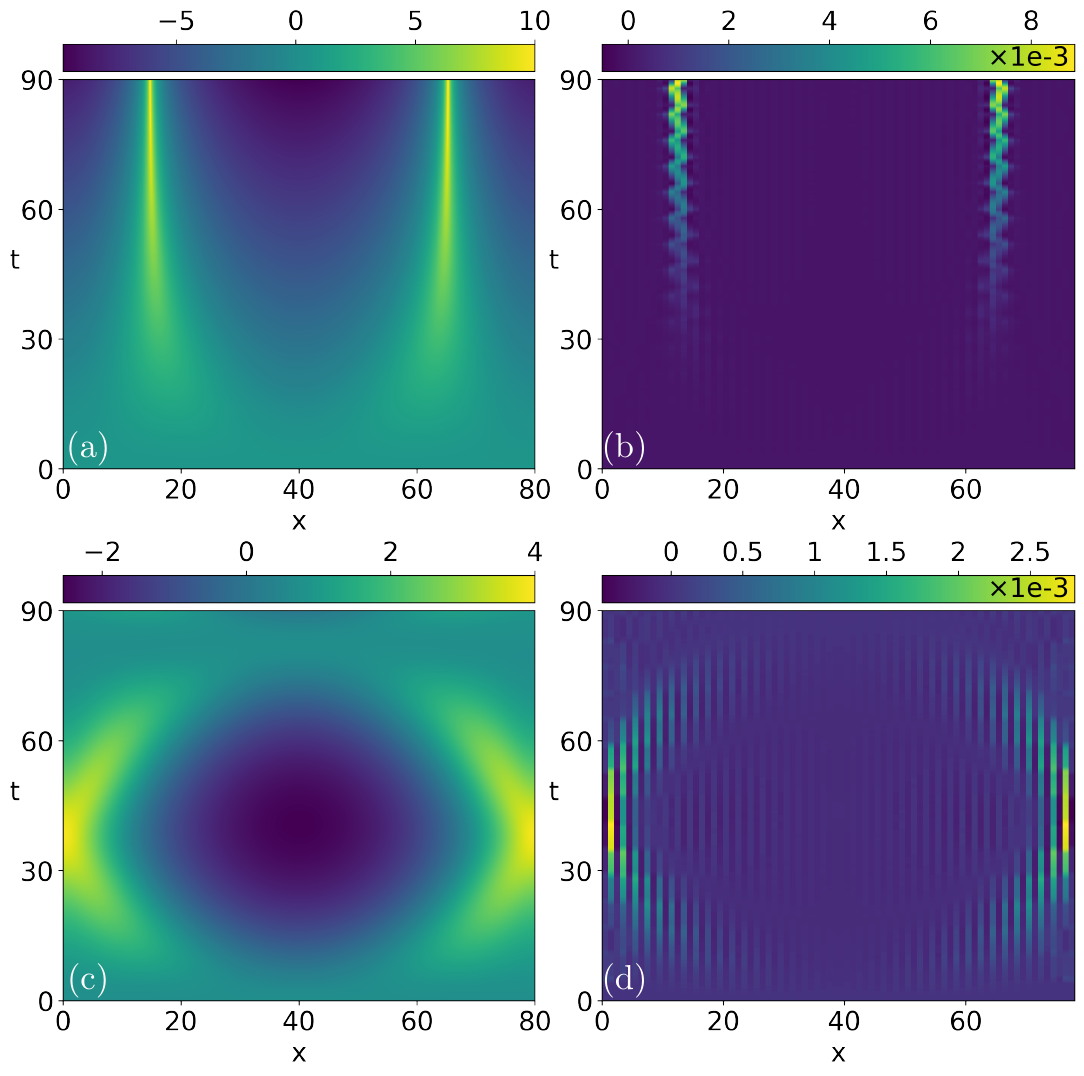}
	\caption{\label{fig:energy} Evolution of energy density $\mathcal E(x,t)$ in both phases, comparing CFT computation with numerical simulations ($L=80$, $T_1=4$, $\Delta=0.5$). In the heating phase (a and b, $T_0=-2$) the energy accumulates indefinitely at the black hole singularities with a time-scale $\Theta_\mathrm{H}^{-1}$. In the non-heating phase (c and d, $T_0=2$) the energy oscillates in the whole chain with pseudo periodicity $T_E$. }
\end{figure}

\textit{Energy propagation} --- The time evolution of the energy density $\mathcal E(x,t) = \langle T_{00}(x,t) \rangle$ provides yet another remarkable validation of the CFT description of the Floquet-SSD XXZ model. For nontrivial energy dynamics, we  choose the ground state $|G\rangle$ of the open chain as the initial state, {because for the periodic chain $\mathcal E(x,0) \equiv 0$ in the ground state}.  The discussion is also applicable to other choices of initial states, such as excited states of the periodic chain. The computation of the time evolution of the energy is similar to the two-point function above, except that $T_{00}$ is not primary and boundaries cannot be neglected. Using  boundary CFT techniques, we obtain~\cite{supmat}
\begin{equation}
	\mathcal E(x,t) = \alpha \left[ \Big(\dfrac{\partial\tilde{z}_n}{\partial z} \Big)^{2} \dfrac{z^2}{\tilde{z}_n^2} + \Big(\dfrac{\partial \bar{\tilde z}_n}{\partial \bar z} \Big)^{2} \dfrac{\bar{z}^2}{\bar{\tilde z}_n^2} \right] -\dfrac{\pi^2 c}{3L^2},
\end{equation}
where $\alpha = \tfrac{c}{32 } (\tfrac{2\pi}{L})^2$, $c$ is the central charge of the theory and $\tilde{z}_n$ is given by Eq.~\eqref{eq:tildez}.  In Fig.~\ref{fig:energy}, we show a comparison between the analytically and numerically obtained energy densities.
 In the non-heating phase, the energy oscillates in time with pseudo-periodicity $T_E$, while in the heating phase, energy accumulates indefinitely at the two horizons $x_c$ and $L-x_c$. Away from these two points, $\mathcal E\to 0$ as $t\rightarrow \infty$, whereas $\mathcal E(x_c,t) \sim \ee^{4\pi\Theta_\mathrm{H} t}$. We see that  heating is  spatially non-uniform and occurs on the time scale $\Theta_\mathrm{H}^{-1}$. 

Finally, the curved space-time description paradoxically suggests the existence of a well defined Floquet effective Hamiltonian  $\HH_\mathrm{eff} = \int \dd x v(x)  T_{00}(x)$ in \emph{both} phases.  Note that  $v=\sqrt{g}$ is the velocity profile appearing in Eq.~\eqref{eq:metric} (the time-reversal symmetric case where $h=0$). We infer~\cite{supmat}
\begin{equation}
  \HH_\mathrm{eff} = \alpha \left[L_0 - \tfrac{\beta}{2} (L_1 + L_{-1}) + {\bar L}_0 - \tfrac{\beta}{2} ({\bar L}_1 + {\bar L}_{-1})\right],
  \end{equation} 
 where  $\beta^{-1}=\cos(\tfrac{\pi T_0}{L})+\frac{L}{\pi T_1}\sin(\tfrac{\pi T_0}{L})$. In the non-heating phase, $|\beta|<1$ and $\HH_\mathrm{eff}$ is related to the uniform theory $H_0$ by an $SL(2,\mathbb R)$ transformation~\cite{supmat}.  At the transition, $\beta=1$ and we recover the SSD Hamiltonian. In the heating phase, $|\beta| >1$  and the relation with $H_0$ instead requires $SL(2,\mathbb C)$, leading to drastically different behavior. In particular, we find that $\HH_\mathrm{eff}$ is unbounded from below in the heating phase as can be seen from the (co)adjoint orbits of $SL(2,\mathbb R)$ \cite{witten1988coadjoint}. A simpler  way to see this is through conformal quantum mechanics \cite{Tada:2017wul,Tada:2019rls}: any Hamlitonian of the form $a L_1 + b L_0 + c L_{-1}$ has a classical counterpart $H = \tfrac{p^2}{2} + V(q)$ with $V(q) = \tfrac{1}{2} \tfrac{1}{q^2} + \tfrac{c^{(2)}}{8}q^2$ with $c^{(2)} = b^2-4a c$ the quadratic Casimir invariant. In our case $c^{(2)} = \alpha^2(1-\beta^2)$ so that $V$ is bounded in the non-heating phase and unbounded in the heating phase.   
Note that in the latter case,  the expression for the density of $\HH_\mathrm{eff}$ bears  marked similarities to   the one of an entanglement Hamiltonian of a subsystem of size $[x_c,L-x_c]$~\cite{Cardy:2016fqc}.

\textit{Discussion} --- The Floquet drive alternating between a uniform and a SSD CFT provides an exactly solvable non-equilibrium system with a rich phase diagram. We present exact analytical results describing the propagation of  excitations  as well as energy density in the system.   Dynamics in the heating phase are understood by analogy to two black-holes singularities and null geodesics in a curved but stationary space-time geometry.
Beyond the timescale fixed by the inverse Hawking temperature, excitations are  absorbed by the  black holes along   with an  accumulation of energy at these singularities. We demonstrate numerically that  the CFT   provides a surprisingly robust  description of  driven  critical spin chains. Cold atomic gases are a promising platform  to realize  such deformed Hamiltonians~\cite{Gross995,monika-unpub}, thereby  opening up the possibility for experimental observation of emergent black hole dynamics in $(1+1)$-dimensional quantum systems.

{\it Note added} --- We note that Ref.~{\onlinecite{Fan_2019}}  also discussed  similar results pertaining to accumulation of energy at two points while  this manuscript was in preparation.

{\it Acknowledgement} --- 
AT thanks Xueda Wen and TN thanks Koji Hashimoto for very helpful discussions. This project has received funding from the European Research Council (ERC) under the European Union's Horizon 2020 research and innovation program (ERC-StG-Neupert-757867-PARATOP and Marie Sklodowska-Curie grant agreement No 701647).

\bibliography{EmergentBlackHoles}

\clearpage
\begin{widetext}

\setcounter{equation}{0}
\renewcommand\theequation{S\arabic{equation}}
\section*{Supplementary Material}

\subsection{Dynamical Two-point function} 
\noindent In this section we compute the dynamical two-point function defined as $F(x,t;x_0,0) \equiv \langle \phi(x,t)\phi(x_0,0) \rangle$ corresponding to a primary field $\phi$ of conformal dimension $h$. The time evolution of the primary is governed by the Floquet Hamiltonian $\HH_\mathrm{F}(t)$ defined in the main text. We closely follow the strategy employed in \cite{Wen:2018agb} wherein the time evolution of the entanglement entropy for a system driven by $\HH_\mathrm{F}(t)$ was computed. Within this setup, we work in imaginary time $\tau$, and introduce Euclidean coordinates $\omega=\tau+\ii x$. Before getting to the computation for an $n$-cycle drive, we describe the 1-cycle drive as a warm-up. The two- point function is 
\begin{equation}
F(x,\tau;x_0,0)=\langle\ee^{\tau_1\HH_{\text{SSD}}} \ee^{\tau_0\HH_{0}}\phi(\omega_1,\bar{\omega}_1)\ee^{-\tau_0\HH_{0}}\ee^{-\tau_1\HH_{\text{SSD}}}\phi(\omega_0,\bar{\omega}_0) \rangle,
\end{equation}
where $\omega_1=0+\ii x$, $\omega_0=0+\ii x_0$ and $\tau=\tau_0+\tau_1$. $\HH_{\text{SSD}}$ and $\HH_{0}$ are the SSD and uniform Hamiltonian described in the main text. Next, under the conformal mapping $z=\exp\left\{\frac{2\pi\omega}{L}\right\}$,  the two-point function transforms as  
\begin{equation}
F(x,\tau;x_0,0)=\left(\frac{2\pi}{L}\right)^{4h}\langle\ee^{\tau_1\HH_{\text{SSD}}} \ee^{\tau_0\HH_{0}}\phi(z_1,\bar{z}_1)\ee^{-\tau_0\HH_{0}}\ee^{-\tau_1\HH_{\text{SSD}}}\phi(z_0,\bar{z}_0)\rangle.
\end{equation}
To compute the time evolution with $\HH_\text{SSD}$ in the complex plane, we introduce the so-called M\"obius Hamiltonian \cite{Okunishi:2016zat}
\begin{equation}
\mathcal{H}_{\text{M\"ob}(\theta)}=L_0-\frac{\tanh(2\theta)}{2}(L_1+L_{-1})+\overline{L}_0-\frac{\tanh(2\theta)}{2}(\overline{L}_1+\overline{L}_{-1}),
\label{mobham}
\end{equation}
defined for $\theta\in\mathbb{R}^+$. Interestingly, there exists an $SL(2,\mathbb{R})$ transformation mapping the M\"obius Hamiltonian to a uniform Hamiltonian. Such mapping is explicitly given by
\begin{equation}
\hat{z}=f(z)=\frac{-\cosh(\theta)z+\sinh(\theta)}{\sinh(\theta)z-\cosh(\theta)}.
\label{mobgood}
\end{equation}
In the $\hat{z}$-coordinates, $\mathcal{H}_{\text{M\"ob}
	(\theta)}\propto \frac{2\pi}{L\cosh(2\theta)}(L_0+\overline{L}_0)$.
Thus the time evolution with $\HH_{\text{M\"ob}(\theta)}$ for a time $\tau$ in the $\hat{z}$-coordinates is a simple dilation by a factor $\lambda = \exp\left\{\frac{2\pi \tau}{L\cosh{2\theta}}\right\}$. Then going back to the original coordinates, the whole time evolution with $\HH_{\text{M\"ob}(\theta)}$ amounts to a simple change of coordinates $z^{\text{new}}_{\theta}(z)=f^{-1}\left(\lambda f(z)\right)$ (in the following of the text we often leave the $z$ dependence of the conformal mappings implicit):
\begin{equation}
z^{\text{new}}_{\theta}(z)=\frac{\left[(1-\lambda)\cosh(2\theta)-(\lambda+1)\right]z+(\lambda-1)\sinh(2\theta)}{(1-\lambda)\sinh(2\theta)z+\left[(\lambda-1)\cosh(2\theta)-(\lambda+1)\right]}.
\label{znew}
\end{equation}
The Hamiltonian $\mathcal{H}_0$ and $\mathcal{H}_{\text{SSD}}$ can be seen as two different limits of the interpolating Hamiltomian $\HH_{\text{M\"ob}(\theta)}$. Indeed, $\HH_0=\HH_{\text{M\"ob}(0)}$ and $\HH_{\text{SSD}}=\HH_{\text{M\"ob}(\theta\rightarrow\infty)}$. From this observation, it may be deduced that one can first evaluate $e^{\tau_0 \HH_{0}}\phi(\omega,\bar{\omega})e^{-\tau_0 \HH_{0}}$ by applying the method in the case $\theta=0$.
\begin{equation}
e^{\tau_0 \HH_{0}}\phi(\omega_1,\bar{\omega}_1)e^{-\tau_0 \HH_{0}}=\left(\frac{2\pi}{L}\right)^{2h}\left[\frac{\partial z^{\text{new}}_{\theta=0}}{\partial z}\bigg|_{z_1}\frac{\partial \bar{z}^{\text{new}}_{\theta=0}}{\partial \bar{z}}\bigg|_{\bar{z}_1}\right]^{h}\phi\left(z^{\text{new}}_{\theta=0}(z_1),\bar{z}^{\text{new}}_{\theta=0}(z_1)\right).
\end{equation}
By looking at the expression for $z^{\text{new}}_{\theta}(z)$ in equation \eqref{znew}, we get $z^{\text{new}}_{\theta=0}(z)=\lambda z$, which is a dilatation in the $z$ plane, as expected for the uniform Hamiltonian $\HH_0$. Next, we need to evaluate 
\begin{align}
\ee^{\tau_1 \HH_{\text{SSD}} }(\ee^{\tau_0 \HH_{0}}\phi(z_1,\bar{z}_1)\ee^{-\tau_0 \HH_{0}})\ee^{-\tau_1 \HH_{\text{SSD}}}\propto \ee^{\tau_1 \HH_{\text{SSD}} }\phi(\lambda z_1,\lambda \bar{z}_1)\ee^{-\tau_1 \HH_{\text{SSD}}},
\end{align}
which can be obtained by using expression of $z^{\text{new}}_{\theta}$ in the limit $\theta\rightarrow\infty$. This just amounts to going to the coordinates $\tilde{z}_1$, defined as 
\begin{equation}
\tilde{z}_1=\lim_{\theta\rightarrow \infty}z^{\text{new}}_{\theta}(\lambda z)= \frac{(1+\frac{\pi\tau_1}{L})e^{\frac{2\pi\tau_0}{L}}z-\frac{\pi\tau_1}{L}}{\frac{\pi \tau_1}{L}e^{\frac{2\pi\tau_0}{L}}z+(1-\frac{\pi\tau_1}{L})}.
\label{1-cycle}
\end{equation}
Hence, $\tilde{z}_1$ is once again related to $z$ by a M\"obius transformation, as expected because it is the obtained via a composition of two (invertible) M\"obius transformations. Consequently the time evolution $\ee^{\tau_1 \HH_{\text{SSD}} }\ee^{\tau_0 \HH_{0}}\phi(z,\bar{z})\ee^{-\tau_0 \HH_{0}}\ee^{-\tau_1 \HH_{\text{SSD}}}$ for a 1-cycle drive of any primary field of a CFT can be reduced to a normalized M\"obius transformation
\begin{equation}
\tilde{z}_1=\frac{az+b}{cz+d},
\label{normalized}
\end{equation}
with
\[
\begin{cases}
a=(1+\frac{\pi\tau_1}{L})e^{\frac{\pi\tau_0}{L}}, \\

b=-\frac{\pi\tau_1}{L}e^{-\frac{\pi\tau_0}{L}}, \\

c=\frac{\pi \tau_1}{L}e^{\frac{\pi\tau_0}{L}},\\

d=(1-\frac{\pi\tau_1}{L})e^{-\frac{\pi\tau_0}{L}}.

\end{cases}
\]
Explicitly the two-point function at different times for a 1-cycle drive is
\begin{equation}
\langle\ee^{\tau_1\HH_{\text{SSD}}} \ee^{\tau_0\HH_{0}}\phi(\omega_1,\bar{\omega}_1)\ee^{-\tau_0\HH_{0}}\ee^{-\tau_1\HH_{\text{SSD}}}\phi(\omega_0,\bar{\omega}_0) \rangle=\left(\frac{2\pi}{L}\right)^{4h}\left[\frac{\partial \tilde{z}_{1}}{\partial z}\bigg|_{z_1}\frac{\partial \bar{\tilde{z}}_1}{\partial \bar{z}}\bigg|_{\bar{z}_1}\right]^{h}\langle\phi(\tilde{z}_{1},\tilde{z}_{1})\phi(z_0,\bar{z}_0)\rangle.
\label{uncycle}
\end{equation}
Therefore we learnt that the time evolution of any primary field during a one cycle of this Floquet drive between $\HH_0$ and $\HH_{\text{SSD}}$ only amounts to a conformal transformation, as seen in \eqref{uncycle}. The main task now is to find how to generalize this result to the full Floquet drive with $n$ cycles. Clearly, the $n$-cycle Floquet time evolution will just amount to a change of coordinates to $\tilde{z}_n$, defined as
\begin{equation}
\tilde{z}_n(z)=\underbrace{(\tilde{z}_1\circ...\circ \tilde{z}_1)}_{n\text{ times}}(z).
\label{eq:composing_Mobius}
\end{equation}
This means that increasing the number of cycles only amounts to composing the 1-cycle transformation with itself.\\
The $n$-cycle M\"obius transformation can be computed by writing the 1-cycle M\"obius transformation in its so-called normal form. Introducing the two fixed-points $\gamma_1$, $\gamma_2$, and the multiplier $\eta$,
\begin{equation}
\begin{cases}
\gamma_1=\frac{a-d-\sqrt{(a-d)^2+4bc}}{2c},  \\

\gamma_2=\frac{a-d+\sqrt{(a-d)^2+4bc}}{2c},  \\

\eta=\frac{(a+d)+\sqrt{(a-d)^2+4bc}}{a+d-\sqrt{(a-d)^2+4bc}}.
\end{cases}
\end{equation}
The normal form of $\tilde{z}_1$ is then
\begin{equation}
\frac{ \tilde{z}_1-\gamma_1}{\tilde{z}_1-\gamma_2}=\eta\frac{z-\gamma_1}{z-\gamma_2}.
\label{normalform}
\end{equation}
It can be shown that in normal form the $n$-cycle evolution simply amounts to
\begin{equation}
\frac{\tilde{z}_n-\gamma_1}{\tilde{z}_n-\gamma_2}=\eta^n\frac{z-\gamma_1}{z-\gamma_2}.
\label{normalform2}
\end{equation}
Then all the stroboscopic time evolution is encoded in the M\"obius multiplier $\eta$. This defines different phases, classified by the trace squared of the $1$-cycle transformation \cite{Wen:2018agb}: 
\begin{align}
\text{Tr}^2\begin{pmatrix}
a&b \\
c&d
\end{pmatrix}=4(1-\Delta).
\end{align}
Indeed if $\Delta>0$ the associated transformation is elliptic and $\eta$ is a phase: the system does not heat. If $\Delta<0$ the associated transformation is hyperbolic and $\eta$ is a positive number: the system heats. $\Delta=0$ corresponds to a parabolic M\"obius transformation, $\eta=1$ and the system is at the phase transition. After analytic continuation, $\Delta$ is written as
\begin{equation}
\Delta=\left[1-\left(\frac{\pi T_1}{L}\right)^2\right]\sin^2\left(\frac{\pi T_0}{L}\right)+\frac{\pi T_1}{L}\sin\left(\frac{2\pi T_0}{L}\right).
\label{delta}
\end{equation}
The $n$-cycles M\"obius transformation can be explicitly written in terms of the parameters of the system as equation \eqref{eq:tildez_app},
\begin{equation}
\tilde{z}_n=\frac{\mathfrak{a} z+\mathfrak{b}}{\mathfrak{c} z +\mathfrak{d}},
\label{eq:tildez_app}
\end{equation}
with:
\[
\begin{cases}
\mathfrak{a}=\gamma_1-\eta^n\gamma_2, \\

\mathfrak{b}=(\eta^n-1)\gamma_1\gamma_2, \\

\mathfrak{c}=1-\eta^n,\\

\mathfrak{d}=\gamma_1\eta^n-\gamma_2.

\end{cases}
\]
Then the stroboscopic time evolution $t=n(T_0+T_1)$ of any primary field $\phi$ can be computed by using this conformal transformation. We stress the fact that here the time evolution is stroboscopic in order to get an analytic handle on the long time dynamics. However, by sacrificing some analytic succinctness we can actually access the full continuous time evolution.
The two-point function at different times is directly obtained with equation \eqref{eq:two_poin},
\begin{equation}\label{eq:two_poin}
\langle \phi(x,t) \phi(x_0,0) \rangle = \left(\frac{2\pi}{L}\right)^{4h}\left[\frac{\partial \tilde{z}_{n}}{\partial z}\bigg|_{z_1}\frac{\partial \bar{\tilde{z}}_n}{\partial \bar{z}}\bigg|_{\bar{z}_1}\right]^{h}\langle \phi(\tilde{z}_n, \bar{\tilde{z}}_n) \phi(\tilde z_0, \bar{\tilde z}_0) \rangle.
\end{equation}
The correlator $\langle \phi(\tilde{z}_n, \bar{\tilde{z}}_n) \phi(\tilde z_0, \bar{\tilde z}_0) \rangle$ can either be computed within the ground state of $\mathcal{H}_0$ with open boundary conditions $|G\rangle$, or the $SL(2,\mathbb{C})$ invariant vacuum $|0\rangle$ of the periodic chain. As $|0\rangle$ is an eigenstate of $\HH_{\text{SSD}}$, the Floquet dynamics should be trivial when computing correlation functions at equal times, as the SSD time evolution is just a phase. However for dynamical two-point functions $\langle 0|e^{\ii \HH_{\text{SSD}}t}\phi(x,0)e^{-\ii \HH_{\text{SSD}}t}\phi(x_0,0)|0\rangle$ the result should not be trivial as $|\Phi\rangle\equiv\phi(x_0,0)|0\rangle$ is not an eigenstate of $\HH_{\text{SSD}}$ in general. Therefore this choice for the computation of $F(x,t;x_0,0)$ is legitimate. In the case of open boundary conditions, we need to use the mapping $z\rightarrow\sqrt{z}$ to map the complex plane with a slit to the upper-half plane, and then evaluate the two point function in the upper-half plane~\cite{Calabrese:2007rg}. This introduces some complications regarding branch cuts of the square root mapping. For simplicity, we choose the periodic case, where
\begin{equation}
\langle 0|\phi(\tilde{z}_n, \bar{\tilde{z}}_n) \phi(\tilde z_0, \bar{\tilde z}_0) |0\rangle \propto \frac{1}{(z_0-\tilde{z}_n)^{2h}}\frac{1}{(\bar{z}_0-\bar{\tilde{z}}_n)^{2h}}.
\end{equation}
This leads to the final formula for the two-point function at different times for $n$-cycles
\begin{equation}
\langle 0|\phi(x,t) \phi(x_0,0) |0\rangle = \left(\frac{2\pi}{L}\right)^{4h}\left[\frac{\partial \tilde{z}_{n}}{\partial z}\bigg|_{z_1}\frac{\partial \bar{\tilde{z}}_n}{\partial \bar{z}}\bigg|_{\bar{z}_1}\right]^{h}\frac{1}{(z_0-\tilde{z}_n)^{2h}}\frac{1}{(\bar{z}_0-\bar{\tilde{z}}_n)^{2h}}.
\end{equation}
It can further be shown that the derivative term simplifies to
\begin{equation}
\frac{\partial \tilde{z}_n}{\partial z}\bigg|_{z_1}\frac{\partial \bar{ \tilde{z}}_n}{\partial \bar{z}}\bigg|_{\bar{z}_1} = \frac{(\mathfrak{a}\mathfrak{d}-\mathfrak{b}\mathfrak{c})^2}{(\mathfrak{c}^2+\mathfrak{d}^2+2\mathfrak{c}\mathfrak{d}\cos\left(\frac{2\pi x}{L}\right))^2}.
\label{derideri}
\end{equation}
In the heating phase, $\eta$ is a real positive number, such that $\eta^n$ tends either to $0$ or $\infty$ depending on the sign of $\eta-1$, corresponding to $\tilde{z}_n$ converging either to $\gamma_1$ or $\gamma_2$. Then $\langle \phi(\tilde{z}_n,\bar{\tilde{z}}_n)\phi(z_0,\bar{z}_0)\rangle$ tends to a constant, and the derivative term \eqref{derideri} is exponentially suppressed for every $x\notin\{x_c,L-x_c\}$, with $x_c$ defined by the fixed points: $\gamma_{1/2}=e^{2\pi x_c/L}$, where $\gamma_{1/2}$ corresponds to $\gamma_{2}$ if $\tilde{z}_n$ converges to $\gamma_1$, and vice-versa. This can be seen explicitly in equation \eqref{limitlongtime},
\begin{equation}
\frac{\partial \tilde{z}_n}{\partial z}\bigg|_{z_1}\frac{\partial \bar{\tilde{z}}_n}{\partial \bar{z}}\bigg|_{\bar{z}_1}= \frac{(\gamma_1-\gamma_2)^4}{\left(\eta^{-n}(1+\gamma_2^2-2\gamma_2\cos{\frac{2\pi x}{L}})+2((\gamma_1+\gamma_2)\cos{\frac{2\pi x}{L}}-1-\gamma_1\gamma_2)+\eta^{n}(1+\gamma_1^2-2\gamma_1\cos{\frac{2\pi x}{L}})\right)^2}.
\label{limitlongtime}
\end{equation}
The expression \eqref{limitlongtime} has a two poles, either in $x_c=\frac{L}{2\pi}\arccos\left(\frac{1+\gamma_2^2}{2\gamma_2}\right)$ and $L-x_c$ if $\lim_{n\rightarrow\infty}\eta^n\rightarrow0$, or in $x_c=\frac{L}{2\pi}\arccos\left(\frac{1+\gamma_1^2}{2\gamma_1}\right)$ and $L-x_c$ if $\lim_{n\rightarrow\infty}\eta^n\rightarrow\infty$. Therefore $F(x,t;x_0,0)$ remains finite even at very long times at these two points, for any choice of $x_0$. The interpretation is that even after a very long number of driving cycles, the excitations will always arrive at one of these two-points at stroboscopic times. Therefore these particular points, only defined with the choice of $T_0/L$ and $T_1/L$, act as attractors for the excitations, which will be better understood within a stroboscopic black-hole picture in an effective space-time in the next section. In the non-heating phase, $\eta$ is a phase, and then after analytic continuation the periodicity reads $T_E=2\pi\frac{ (T_0+T_1)}{|\arg(\eta)|}$. The excitations are then propagating periodically with $T_E$. However if $T_E/(T_0+T_1) \notin \mathbb Q$, the system is pseudo-periodic, as the two-point function is only defined at stroboscopic times.\\

\subsection{Effective curved space-time} 
\noindent The two-point function at different times $F(x,t;x_0,0)$ enables us to get the light-cone propagation of the gapless excitations. For homogeneous Luttinger liquids, the excitations are following straight lines in space-time. However for inhomogeneous Luttinger liquid with spatial deformation $v(x)$, they are following curves, which are nothing more than light-like geodesics in an effective curved space-time specified by the metric $\dd s^2=\dd x^2+v(x)^2\dd \tau^2$ \cite{Dubail_2017}. In the case of the sine-square deformation, the metric is $\dd s^2=\dd x^2+\sin^2\left(\frac{\pi x}{L}\right)\dd \tau^2$. Thus, the null geodesics are simply given by the light-like condition $\dd s^2=0$, giving the propagation of the excitations starting at $x_0$: $x_{\pm}(t)=\frac{L}{\pi}\cot^{-1}{\left(\pm \frac{2\pi}{L} t +\cot{\frac{\pi x_0}{L}}\right)}$. Therefore it is clear that the excitations never reach the boundaries of the system in this case, as their local group velocity goes to $0$ at the edges.

\medskip \noindent We now derive the effective space-time metric for the Floquet drive defined at stroboscopic times.
We are interested in finding some coordinates $\tilde{z}$ in which the effective metric describing the $n$-cycle Floquet drive is conformally flat, and then going back to the original coordinates $(x,\tau)$ to get the expression of the metric. Such coordinates are called isothermal coordinates and always exist in $(1+1)$-dimensional space-times. For the Floquet drive, they are given by the effective M\"obius transformation \eqref{eq:tildez_app}, so that the metric reads
\begin{equation}
\dd s^2=\dd \tilde{z}_n\dd \bar{\tilde{z}}_n.
\end{equation}
Introducing the real and imaginary parts of $\tilde{z}$
\[
\begin{cases}
\tilde{u}_n(x,\tau)=\text{Re}(\tilde{z}_n)=\frac{\mathfrak{a}\mathfrak{c} + \mathfrak{b}\mathfrak{d} +(\mathfrak{a}\mathfrak{d}+\mathfrak{b}\mathfrak{c})\cos{\left(\frac{2\pi x}{L}\right)}}{\mathfrak{c}^2+\mathfrak{d}^2+2\mathfrak{c}\mathfrak{d}\cos{\left(\frac{2\pi x}{L}\right)}},\\
\tilde{v}_n(x,\tau)=\text{Im}(\tilde{z}_n)=\frac{(\mathfrak{a}\mathfrak{d}-\mathfrak{b}\mathfrak{c})\sin{\left(\frac{2\pi x}{L}\right)}}{\mathfrak{c}^2+\mathfrak{d}^2+2\mathfrak{c}\mathfrak{d}\cos{\left(\frac{2\pi x}{L}\right)}}.
\end{cases}
\]
The effective metric reads $\dd s^2=\dd \tilde{u}_n^2+\dd \tilde{v}_n^2$.
It is now straightforward to apply the change to ($x$,$\tau$) coordinates.
After some computations, the metric takes the familiar form, after analytic continuation:
\begin{equation*}
\dd s^2= \ee^{2\sigma(x,\tau)}\left(\dd x^2 - g(x)\dd \tau^2 + 2 h(x) \dd x \dd t \right).
\label{souri}
\end{equation*}
The value we find for $g(x)$ and $h(x)$ are then given by equations \eqref{efe} and \eqref{timerev}
\begin{equation}
g(x)= \zeta^2\prod_{i=1}^2\left[1+\gamma_i^2-2\gamma_i\cos{\left(\frac{2\pi x}{L}\right)}\right],
\label{efe}
\end{equation}
\begin{equation}
h(x)= \ii\zeta (\gamma_1\gamma_2 - 1)
\sin\left(\frac{2\pi x}{L}\right),
\label{timerev}
\end{equation}
where $\zeta= -\frac{L}{2\pi \ii}\frac{1}{(T_0 + T_1)}\frac{\log(
	\eta)}{(\gamma_1 - \gamma_2)}$, and as before $\eta$ is the multiplier of the M\"obius transformation, which is a complex exponential in the non-heating phase and a real exponential in the heating phase, and $\gamma_1$, $\gamma_2$ are the two fixed-points of the 1-cycle M\"obius transformation. After analytic continuation, both $g(x)$ and $h(x)$ are real-valued functions.\\
The Weyl prefactor $\ee^{2\sigma(x,\tau)}$ is a positive number before analytic continuation,
\begin{equation}
\ee^{2\sigma(x,\tau)}=\frac{4\pi^2}{L^2}\frac{\eta^{2n}(\gamma_1-\gamma_2)^4}{\left(1 + \eta^{2n} (1 + \gamma_1^2) + \gamma_2^2 - 
	2\eta^n (1 + \gamma_1 \gamma_2) - 
	2 (-1 + \eta^n) (\eta^n \gamma_1 - 
	\gamma_2)\cos(\frac{2\pi x}{L})\right)^2}.
\end{equation}
Inverting the Weyl transformation, the metric is finally given by
\begin{equation}
\dd s^2= \dd x^2-g(x) \dd t^2 +2h(x)\dd x\dd t.
\label{eq:metricnontrv}
\end{equation}
The null geodesics of this $(1+1)$d space-time are uniquely determined by the condition $\dd s^2=0$. Thus they are the solutions of the equation $ 2h(x(t))\dot{x}(t)+\dot{x}^2(t)-g(x(t))=0$: 
\begin{equation}
\pm t(x)= \int_{x_0}^{x}dx'\frac{1}{\sqrt{h(x')^2+g(x')}\mp h(x')}.
\end{equation}
Then the local group velocity of the excitations is $v(x) = h(x) \mp \sqrt{h(x)^2+g(x)}$, where the sign corresponds to chiral and anti-chiral excitations. The expression \eqref{eq:metricnontrv} is not time-reversal invariant because of the off-diagonal term $h(x)$. Only if $\gamma_1\gamma_2=1$, $h(x)=0$ and the system is time-reversal invariant. This can be fulfilled by starting the drive in a symmetric way. For concreteness, we shift the origin of time by $\frac{T_0}{2}$
\begin{align}
\HH_\mathrm{F}(t)=
\begin{cases}
\HH_0 & 0<t<\frac{T_0}{2},\\              
\HH_{\text{SSD}}&\frac{T_0}{2}<t<\frac{T_0}{2}+T_1,\\
\HH_0&\frac{T_0}{2}+T_1<t<\frac{3T_0}{2}+T_1,\\
\text{etc.}
\end{cases}
\end{align}
The associated 1-cycle M\"obius transformation is therefore given by the equation
\begin{equation}
\tilde{z}_1=\frac{\left(1+\frac{\pi \tau_1}{L}\right)\ee^{\frac{\pi\tau_0}{L}}z-\frac{\pi\tau_1}{L}}{\frac{\pi\tau_1}{L}z+\left(1-\frac{\pi\tau_1}{L}\right)\ee^{-\frac{\pi\tau_0}{L}}}.
\label{link}
\end{equation}
It is interesting to compare \eqref{link} and \eqref{normalized}.
The coefficients $a$ and $d$ are the same as in the non symmetric case, but not $b$ and $c$. Furthermore $bc$ is unchanged. Thus, looking at the definitions of the fixed points and the multiplier, this only redefines the denominators of $\gamma_1$ and $\gamma_2$, and keeps the multiplier $\eta$ invariant. It can then be shown that $\gamma_1\gamma_2=1$. Furthermore in the heating phase $|\gamma_1|=|\gamma_2|=1$, therefore in the time-reversal symmetric situation $\gamma_1=\gamma_2^*$, whereas in the non-heating phase, $\gamma_1$ and $\gamma_2$ are both real and inverse of each other. In this case, the metric is simplified to $\dd s^2=\dd x^2-g(x)\dd t^2$. Applying the time-reversal condition, one finds that
\begin{equation}
v(x)=[g(x)]^{1/2}= \frac{1}{2\pi \ii}\frac{L\log(\eta)}{(T_0+T_1)} \frac{\left(1+\gamma_1^2-2\gamma_1\cos{\frac{2\pi x}{L}}\right)}{\gamma_1^2-1}.
\label{vitevite}
\end{equation}
This is the effective velocity of the excitations. In the heating phase their local group velocity goes to 0 at two points, which are found to be $x_c=\frac{L}{2\pi}\arccos\left(\frac{1+\gamma_1^2}{2\gamma_1}\right)$ and $L-x_c$. After analytic continuation, it can be shown that $x_c= \tfrac{L}{2\pi} \arccos(\cos \tfrac{\pi T_0}{L} + \tfrac{L}{\pi T_1}  \sin \tfrac{\pi T_0}{L})$. Thus $x_c\in [0,\frac{L}{2}]$ in the heating phase, and $x_c$ is a complex number in the non-heating phase, therefore the velocity never goes to $0$. Thus the heating phase at these two points, the velocity of the excitations vanishes, meaning that their wordlines, following null geodesics of the metric, will tend to one of these two points.
We rewrite the metric in the heating phase in terms of the singularity $x_c$. We first notice that as
$\cos\left(\frac{2\pi x_c}{L}\right)=\frac{1}{2}\frac{\gamma_1^2+1}{\gamma_1}$, then
the effective deformation is rewritten directly in terms of the singularity
\begin{equation}
v(x)= A \left(1-\frac{\cos\left(\frac{2\pi x}{L}\right)}{\cos\left(\frac{2\pi x_c}{L}\right)}\right),
\end{equation}
with $A=\frac{1+\gamma_1^2}{\gamma_1^2-1}\frac{1}{2\pi \ii(T_0+T_1)}L\log(\eta)$. Using trigonometric formulae, this leads to the desired form of the velocity
\begin{equation}
v(x)= 2A  \frac{\sin\left[\frac{\pi}{L}(x-x_c)\right]\sin\left[\frac{\pi}{L}(x+x_c)\right]}{\cos\left(\frac{2\pi x_c}{L}\right)},
\label{vitess}
\end{equation}
The effective metric can now be easily expressed in terms of the singularity $x_c$
\begin{equation}
\dd s^2=-4A^2\frac{\sin^2\left[\frac{\pi}{L}(x-x_c)\right]\sin^2\left[\frac{\pi}{L}(x+x_c)\right]}{\cos^2\left(\frac{2\pi x_c}{L}\right)}\dd t^2+\dd x^2.
\label{basbas}
\end{equation}
This form is still hard to interpret in terms of Schwarzschild metric. However, by doing an expansion around one of the two singularities, i.e., around $x_c$ or $L-x_c$, and keeping only the lowest order contribution, only the contribution from one of the singularities should matter and the metric should be simpler. Therefore we expand the expression \eqref{basbas} around $x_c$. At leading order in $x-x_c$, we may simplify 
\begin{align}
\sin^2\left[\frac{\pi}{L}(x-x_c)\right]\sin^2\left[\frac{\pi}{L}(x+x_c)\right]\approx \frac{\pi^2}{L^2}(x-x_c)^2\sin^2\left(\frac{2\pi x_c}{L}\right) + \mathcal{O}((x-x_c)^3).
\end{align}
The metric finally simplifies to
\begin{equation}
\dd s^2= -4A^2\tan^2\left(\frac{2\pi x_c}{L}\right)\frac{\pi^2}{L^2}(x-x_c)^2 \dd t^2+\dd x^2.
\label{rindler}
\end{equation}
This metric is known as the Rindler metric, which describes an accelerated frame transformation of the flat Minkowski metric. Writing $C^2=4A^2\tan^2\left(\frac{2\pi x_c}{L}\right)\frac{\pi^2}{L^2}$, the metric reads $\dd s^2= -C^2 (x-x_c)^2\dd t^2+\dd x^2$.
One can now introduce the following coordinate change: $\frac{C}{2}\left(x-x_c\right)^2=\left(y-x_c\right)$.
In the new coordinates, the metric reads
\begin{equation}
\dd s^2=-2C\left(y-x_c\right)\dd t^2+\frac{1}{2C}\frac{1}{\left(y-x_c\right)}\dd y^2.
\end{equation} 
This is the well-known Schwarzschild metric in $(1+1)$ dimensions. Thus expanding our space-time effective metric around one of the two singularities gives (at leading order) a black hole metric. One can also do that for the second singularity by expanding the metric around $L-x_c$, to get similar results: $\dd s^2=-2C\left[y-(L-x_c)\right]\dd t^2+\frac{1}{2c}\frac{1}{\left[y-(L-x_c)\right]}\dd y^2$.

\medskip \noindent The Hawking temperature $\Theta_H$ can be directly read-off from the expression of the metric as $\Theta_H=\frac{C}{2\pi}$. We can finally use the formula $\tan{\left(\frac{2\pi x_c}{L}\right)}=\frac{1}{\ii}\frac{\gamma_1^2-1}{\gamma_1^2+1}$ to conclude that the Hawking temperature is given by $\Theta_H=\frac{|\log{(\eta)}|}{2\pi(T_0+T_1)}$.\\

\subsection{Effective Hamiltonian}
\noindent Using the effective metric in the time-reversal symmetric case, we deduce that the stroboscopic effective Hamiltonian is $\mathcal{H}_{\text{eff}}=\int_0^L v(x)T_{00}(x)\dd x$. Using the Fourier decomposition of $v(x)$, given by the equation \eqref{vitevite}, and using the definition of the Virasoro generators $L_n=\frac{1}{2\pi \ii}\oint \dd z z^{n+1}T(z)$, we can conclude that the stroboscopic Hamiltonian is
\begin{equation}
\HH_{\text{eff}}=\alpha\left[L_0-\frac{\beta}{2}(L_1+L_{-1})+\overline{L}_0-\frac{\beta}{2}(\overline{L}_1+\overline{L}_{-1})\right],
\end{equation}
where $\alpha=\frac{1+\gamma_1^2}{\gamma_1^2-1}\frac{L}{2\pi \ii(T_0+T_1)}\log{(\eta)}$, $\beta=\frac{2\gamma_1}{1+\gamma_1^2}$, which are real numbers. It can further be shown using the expressions of the fixed-points that: $\beta^{-1}=\cos(\frac{\pi T_0}{L})+\frac{L}{\pi T_1}\sin(\frac{\pi T_0}{L})$. Therefore in the heating phase $|\beta|>1$, and in the non-heating phase $|\beta|<1$. 

\medskip \noindent  In the case $|\beta|<1$, the effective Hamiltonian is simply the M\"obius Hamiltonian \eqref{mobham}. This observation is consistent with the fact that $F(x,t;x_0,0)$ is periodic in the non-heating phase. Indeed, the propagation of the excitations after a quench with the M\"obius Hamiltonian is also periodic, with period $T=\frac{1}{L\cosh(2\theta)}$. Therefore the effective stroboscopic Hamiltonian in the non-heating phase in the time-reversal symmetric case is just an interpolating Hamiltonian between $\HH_0$ and $\HH_{\text{SSD}}$. 
$\HH_{\text{eff}}$ can be further written as the convex combination of the two original Hamiltonians
\begin{equation}
\mathcal{H}_{\text{eff}}=\alpha\left[(1-\beta)\mathcal{H}_0+\beta\mathcal{H}_{\text{SSD}}\right].
\end{equation}
Therefore, for $0<\beta<1$, the effective Hamiltonian interpolates between the uniform and the SSD Hamiltonian, as we already understood through the comparison with the M\"obius Hamiltonian. 

\medskip \noindent For $\beta>1$, the effective Hamiltonian cannot be understood as an interpolation between the two original Hamiltonians, giving rise to the physics of heating. The effective Hamiltonian in the heating phase can be rewritten, using \eqref{vitess} as:
\begin{equation}
\HH_{\text{eff}}=2L\Theta_H\int_0^L\dd x\frac{ \sin\left(\frac{\pi}{L}(x-x_c)\right)\sin\left(\frac{\pi}{L}(x+x_c)\right)}{\sin\left(\frac{2\pi x_c}{L}\right)}T_{00}(x).
\end{equation}
This form of the effective Hamiltonian is reminiscent of the entanglement Hamiltonian $K_A$ for a system of finite size $[0,L]$, introduced in \cite{Cardy:2016fqc}, with subsystem $A=(x_c,L-x_c)$. However here the Hamiltonian density is integrated over the whole chain. For such entanglement Hamiltonian, an effective local temperature can be defined, diverging at $x_c$ and $L-x_c$. This is an indication that energy should be absorbed exponentially at these two points.

\medskip \noindent Finally, looking at the effective deformation $v(x)$ is insightful: in the non-heating phase, we notice that $v(x)$ has no roots. Therefore, it can be thought of as a shifted sine-square, deforming the homogeneous system only smoothly. By going through the phase diagram following the line $T_0=T_1$, this shifted sine-square will simply tend to the usual sine-square. At the phase transition, the effective Hamiltonian is similar to the sine-square deformation, and has roots at $x_c=0$ and $x_c=L$. Then, in the heating phase, the two roots will approach symmetrically to the center of the system, giving rise to a cosine-square deformation at the second phase transition , having a single root at $x_c=\frac{L}{2}$. Therefore, the effective Hamiltonian in the heating and non-heating phase only interpolates between the sine-square and the cosine-square deformations.

\subsection{Energy density} 
\noindent The different phases arising within the Floquet CFT were understood in \cite{Wen:2018agb} by computing the entanglement entropy, which grows linearly in the heating phase and oscillates with period $T_E$ in the non-heating phase. We would like to characterize more precisely these phases by computing the evolution of the energy density $\mathcal{E}(x,t)$ in the system. In particular, we expect to observe an exponential increase of energy in the heating phase precisely at the location of the two singularities $x_c$ and $L-x_c$, whereas the rest of the system should not absorb energy, to agree with our stroboscopic black hole picture. The energy density $\mathcal{E}(x,t)$ under the Floquet drive is defined by 
\begin{equation}
\mathcal{E}(x,t)=\langle \psi(t)|T_{00}(x)|\psi(t)\rangle.
\label{njr}
\end{equation}
As usual $T_{00}$ is the energy density of the uniform CFT. $|\psi(t)\rangle=U(t)|G\rangle$ is the time evolved ground state of the uniform Hamiltonian $H_0$, with open boundary conditions. We chose open boundary conditions in this case, as in the periodic case $\mathcal{E}(x,t)=0$. In Euclidean coordinates, $T_{00}=T(\omega)+\bar{T}(\bar{\omega})$. The strategy is the same as for the two point function at different times: the first step is to map the strip to the complex plane with a slit, using the exponential mapping
\begin{equation}
\langle G|\ee^{\tau \HH_{\text{SSD}}}T(\omega)\ee^{-\tau \HH_{\text{SSD}}}|G\rangle=\left(\frac{\partial z}{\partial \omega}\right)^2 \langle G|\ee^{\tau \HH_{\text{SSD}}}T(z)\ee^{-\tau \HH_{\text{SSD}}}|G\rangle-\frac{\pi^2 c}{6L^2}.
\end{equation}
Then, the usual procedure consists in mapping the complex plane to itself in the M\"obius $\tilde{z}_n$ coordinates, applying the time evolution and transforming back to the $z$ coordinates. The extra terms coming from the Schwarzian derivative vanish because of $SL(2,\mathbb{R})$ invariance
\begin{equation}
\langle G|\ee^{\tau \HH_{\text{SSD}}}T(\omega)\ee^{-\tau \HH_{\text{SSD}}}|G\rangle=\left(\frac{\partial z}{\partial \omega}\right)^2 \left(\frac{\partial \tilde{z}_n}{\partial z}\right)^2 \langle G|T(\tilde{z}_n)|G\rangle-\frac{\pi^2 c}{6L^2}.
\end{equation}
The final step is to evaluate the correlation function $\langle G|T(\tilde{z}_n)|G\rangle$ in a boundary CFT, defined on the complex plane with a slit on the real positive axis. This can be done using a square-root mapping $\sqrt{z}$ to the upper-half plane $\mathbb{H}$. This gives a non-trivial Schwarzian derivative term given by $\{z,\sqrt{z}\}=\dfrac{3}{8z^2}$. The upper-half plane can me mapped to the unit disc with a M\"obius transformation, therefore, due to rotational symmetry, $\langle G|T(\sqrt{\tilde{z}_n})|G\rangle_{\mathbb{H}}=0$ \cite{Calabrese:2009qy}. Finally, only the Schwarzian derivative term of the square root mapping contributes to the energy density, which before analytic continuation reads
\begin{equation}
\mathcal{E}(x,t)= \frac{c}{32}\left[\left(\frac{\partial z}{\partial \omega}\right)^2 \left(\frac{\partial \tilde{z}_n}{\partial z}\right)^2\frac{1}{\tilde{z}_n^2}+\left(\frac{\partial \bar{z}}{\partial \bar{\omega}}\right)^2 \left(\frac{\partial \bar{\tilde{z}}_n}{\partial \bar{z}}\right)^2\frac{1}{\bar{\tilde{z}}_n^2}\right]-\frac{\pi^2 c}{3L^2},
\end{equation}
for stroboscopic times $t=n(T_0+T_1)$. In the heating phase, as $\mathcal{E}(x_c,t)\sim \eta^{-2n}$ at long time because of the derivative term $\frac{\partial \bar{\tilde{z}}_n}{\partial \bar{z}}$, we conclude that $\mathcal E(x_c,t) \sim \ee^{4\pi\Theta_\mathrm{H} t}$, such that the Hawking temperature is really the heating rate. Similarly for the other singularity $L-x_c$, where the energy is also growing exponentially because of the other derivative term $\frac{\partial \tilde{z}_n}{\partial z}$. Therefore the energy density grows exponentially in the heating phase only at the positions of the two black holes, as expected. In the non-heating phase, the energy density oscillates in time with period $T_E=2\pi\frac{ (T_0+T_1)}{|\arg(\eta)|}$.
\end{widetext}
\end{document}